\documentclass[11pt, letterpaper]{article}
\usepackage[colorlinks=true,linkcolor=red,citecolor=blue]{hyperref}
\usepackage{fullpage,paralist}
\usepackage{theorem}
\usepackage{times}
\usepackage{booktabs}
\usepackage{stmaryrd}

\usepackage{mathptmx}

\usepackage[text={6.5in,9in}]{geometry}

\usepackage{verbatim}
\usepackage{amssymb}
\usepackage{amsmath}
\usepackage{ifthen}

\DeclareMathOperator*{\argmin}{arg\,min}

\newtheorem{theorem}{Theorem}[section]

\newtheorem{remark}{Remark}

\newtheorem{corollary}{Corollary}[section]

\newskip\theorempreskipamount
\newskip\theorempostskipamount
\global
\global

\newenvironment{reminder}[1]{\smallskip
\noindent {\bf Reminder of #1  }\em}{}
\newenvironment{proof}{\noindent {\bf Proof.  }}{\hfill$\Box$}
\newenvironment{proofof}[1]{\smallskip
\noindent {\bf Proof of #1.  }}{\hfill$\Box$}

\newtheorem{lemma}{Lemma}[section]
\newtheorem{claim}{Claim}
\def \Z {{\mathbb Z}}
\def \AC {{\sf AC}}
\def \THR {{\sf THR}}

\def \poly {{\text{poly}}}
\def \R {{\mathbb R}}

\def \SYM {{\sf SYM}}
\def \ACC {{\sf ACC}}

\def \N {{\mathbb N}}
\def \F {{\mathbb F}}
\def \eps {{\varepsilon}}

\title{Faster all-pairs shortest paths via circuit complexity\footnote{This is a preliminary version; comments are welcome.}}
\author{Ryan Williams\thanks{Supported by an Alfred P. Sloan Fellowship, a Microsoft Research Faculty Fellowship, a David Morgenthaler II Faculty Fellowship, and NSF CCF-1212372. Any opinions, findings, and conclusions or recommendations
expressed in this material are those of the author(s) and do not necessarily reflect the views of the National Science Foundation.}\\Stanford University}

\begin{document}
\maketitle

\begin{abstract} We present a new randomized method for computing the min-plus product (a.k.a., tropical product) of two $n \times n$ matrices, yielding a faster algorithm for solving the all-pairs shortest path problem (APSP) in dense $n$-node directed graphs with arbitrary edge weights. On the real RAM, where additions and comparisons of reals are unit cost (but all other operations have typical logarithmic cost), the algorithm runs in time \[\frac{n^3}{2^{\Omega(\log n)^{1/2}}}\] and is correct with high probability. On the word RAM, the algorithm runs in $n^3/2^{\Omega(\log n)^{1/2}} + n^{2+o(1)}\log M$ time for edge weights in $([0,M] \cap \Z)\cup\{\infty\}$.
Prior algorithms took either $O(n^3/\log^c n)$ time for various $c \leq 2$, or $O(M^{\alpha}n^{\beta})$ time for various $\alpha > 0$ and $\beta > 2$.

The new algorithm applies a tool from circuit complexity, namely the Razborov-Smolensky polynomials for approximately representing $\AC^0[p]$ circuits, to efficiently reduce a matrix product over the $(\min,+)$ algebra to a relatively small number of rectangular matrix products over $\F_2$, each of which are computable using a particularly efficient method due to Coppersmith. We also give a deterministic version of the algorithm running in $n^3/2^{\log^{\delta} n}$ time for some $\delta > 0$, which utilizes the Yao-Beigel-Tarui translation of $\AC^0[m]$ circuits into ``nice'' depth-two circuits.
\end{abstract}

\thispagestyle{empty}


\newpage
\setcounter{page}{1}

\section{Introduction}

The all-pairs shortest path problem (APSP) and its $O(n^3)$ time solution on $n$-node graphs~\cite{Floyd,Warshall} are standard classics of computer science textbooks. To recall, the input is a weighted adjacency matrix of a graph, and we wish to output a data structure encoding all shortest paths between any pair of vertices---when we query a pair of nodes $(s,t)$, the data structure should reply with the shortest distance from $s$ to $t$ in $\tilde{O}(1)$ time, and a shortest path from $s$ to $t$ in $\tilde{O}(\ell)$ time, where $\ell$ is the number of edges on the path. As the input to the problem may be $\Theta(n^2 \cdot \log M)$ bits (where $M$ bounds the weights), it is natural to wonder if the $O(n^3)$ bound is the best one can hope for.\footnote{It is not obvious that $o(n^3)$-size data structures for APSP should even exist! There are $n^2$ pairs of nodes and their shortest paths may, in principle, be of average length $\Omega(n)$. However, representations of size $\Theta(n^2 \log n)$ do exist, such as the ``successor matrix'' described by Seidel~\cite{Seidel}.} (In fact, Kerr~\cite{Kerr70} proved that in a model where only additions and comparisons of numbers are allowed, $\Omega(n^3)$ operations are required.)

Since the 1970s~\cite{Munro71,Fischer-Meyer71,AhoHopcroftUllman} it has been known that the search for faster algorithms for APSP is equivalent to the search for faster algorithms for the min-plus (or max-plus) matrix product (a.k.a. distance product or tropical matrix multiplication), defined as: \[(A \star B)[i,j] = \min_{k} (A[i,k] + B[k,j]).\] That is, $\min$ plays the role of addition, and $+$ plays the role of multiplication. A $T(n)$-time algorithm exists for this product if and only if there is an $O(T(n))$-time algorithm for APSP.\footnote{Technically speaking, to reconstruct the shortest paths, we also need to compute the product $(A \odot B)[i,j] = \argmin_{k} (A[i,k] + B[k,j])$, which returns (for all $i,j$) some $k$ witnessing the minimum $A[i,k]+B[k,j]$. However, all known distance product algorithms (including ours) have this property.}

Perhaps inspired by the surprising $n^{2.82}$ matrix multiplication algorithm of Strassen~\cite{Strassen} over \emph{rings},\footnote{As $\min$ and $\max$ do not have additive inverses, min-plus algebra and max-plus algebra are not rings, so fast matrix multiplication algorithms do not directly apply to them.} Fredman~\cite{Fredman} initiated the development of $o(n^3)$ time algorithms for APSP. He discovered a \emph{non-uniform} decision tree computing the $n \times n$ min-plus product with depth $O(n^{2.5})$ (but with size $2^{\Theta(n^{2.5})}$). Combining the decision tree with a lookup table technique, he obtained a uniform APSP algorithm running in about $n^3/\log^{1/3} n$ time. Since 1975, many subsequent improvements on Fredman's algorithm have been reported (see Table~\ref{APSPtable}).\footnote{There have also been parallel developments in APSP algorithms on sparse graphs~\cite{Johnson1977,FredmanTarjan,PettieR,Pettie,Chan12} and graphs with small integer weights~\cite{Romani,Pan81,Seidel,Takaoka98,AlonGalilMargalit,ShoshanZwick, Zwick02,GabowSankowski}. The small-weight algorithms are \emph{pseudopolynomial}, running in time $O(M^{\alpha} n^{\beta})$ for various $\alpha > 0$ and various $\beta$ greater than the (ring) matrix multiplication exponent.} However, all these improvements have only saved $\log^{c} n$ factors over Floyd-Warshall: most recently, Chan~\cite{Chan10} and Han and Takaoka~\cite{HanT12} give time bounds of roughly $n^3/\log^2 n$.

The consensus appears to be that the known approaches to general APSP may never save more than small $\poly(\log n)$ factors in the running time. The methods (including Fredman's) use substantial preprocessing, lookup tables, and (sometimes) bit tricks, offloading progressively more complex operations into tables such that these operations can then be executed in constant time, speeding up the algorithm. It is open whether such techniques could even lead to an $n^3/\log^3 n$ time Boolean matrix multiplication (with logical OR as addition), a special case of max-plus product. V.~Vassilevska Williams and the author~\cite{VW10} proved that a large collection of fundamental graph and matrix problems are \emph{subcubic equivalent} to APSP: Either all these problems are solvable in $n^{3-\eps}$ time for some $\eps > 0$ (a.k.a. ``truly subcubic time''), or none of them are. This theory of APSP-hardness has nurtured some pessimism that truly subcubic APSP is possible.

\begin{table*}\label{APSPtable}\centering
\begin{tabular}{@{}lll@{}}
\toprule
Time  &	Author(s) 			&	Year(s) \\ \midrule
$n^3$ & 	Floyd~\cite{Floyd}/Warshall~\cite{Warshall} & 1962/1962\\
$n^3 / \log^{1/3} n$ & Fredman~\cite{Fredman}  &  1975\\
$n^3 / \log^{1/2} n$ & Dobosiewicz~\cite{Dob90}/Takaoka~\cite{Takaoka91} & 1990/1991\\
$n^3 / \log^{5/7} n$ & Han~\cite{Han04} &  2004\\
$n^3 / \log n$ & Takaoka~\cite{Takaoka04}/Zwick~\cite{Zwick04}/Takaoka~\cite{Takaoka05}/Chan~\cite{Chan08} &   2004/2004/2005/2005 \\
$n^3 / \log^{5/4} n$ & Han~\cite{Han08} & 2006\\
$n^3 / \log^2 n$&  Chan~\cite{Chan10}/Han-Takaoka~\cite{HanT12} &  2007/2012\\
$n^3 / 2^{\Omega(\log n)^{1/2}}$ & this paper & \\
\bottomrule
\end{tabular}
\caption{Running times for general APSP, omitting poly(log log n) factors. Years are given by the earliest conference/journal publication. (Table adapted from Chan~\cite{Chan10}.)}
\end{table*}

We counter these doubts with a new algorithm for APSP running faster than $n^3/\log^k n$ time, for every $k$.

\begin{theorem}\label{main} On the word RAM, APSP can be solved in \[n^3/2^{\Omega(\log n)^{1/2}} + n^{2+o(1)}\log M\] time with a Monte Carlo algorithm, on $n$-node graphs with edge weights in $([0,M]\cap \Z)\cup\{\infty\}$. On the real RAM, the $n^{2+o(1)}\log M$ factor may be removed.
\end{theorem}

\begin{remark} A similar $n^{2+o(1)}\log M$ time factor would necessarily appear in a complete running time description of all previous algorithms when implemented on a machine that takes bit complexity into account, such as the word RAM---note the input itself requires $\Omega(n^2 \log M)$ bits to describe in the worst case. In the \emph{real RAM} model, where additions and comparisons of real numbers given in the input are unit cost, but all other operations have typical cost, the algorithm runs in the ``strongly polynomial'' bound of $n^3/2^{\Omega(\log n)^{1/2}}$ time. Most prior algorithms for the general case of APSP also have implementations in the real RAM. (For an extended discussion, see Section 2 of Zwick~\cite{Zwick04}.)
\end{remark}

The key to Theorem~\ref{main} is a new reduction from min-plus (and max-plus) matrix multiplication to (rectangular) matrix multiplication over $\F_2$. To the best of our knowledge, all prior reductions from $(\max,+)$ algebra to (the usual) $(+,\times)$ algebra apply the mapping $a \mapsto x^a$ for some sufficiently large (or sometimes indeterminate) $x$. Under this mapping, $\max$ ``maps to'' $+$ and $+$ ``maps to'' $\times$: $\max\{a,b\}$ can be computed by checking the degree of $x^a + x^b$, and $a+b$ can be computed by $x^a \times x^b = x^{a+b}$. Although this mapping is extremely natural (indeed, it is the starting point for the field of tropical algebra), the computational difficulty with this reduction is that the sizes of numbers increase exponentially.

The new algorithm avoids an exponential blowup by exploiting the fact that $\min$ and addition are \emph{simple} operations from the point of view of Boolean circuit complexity. Namely, these operations are both in $\AC^0$, i.e., they have circuits of constant-depth and polynomial-size over AND/OR/NOT gates of unbounded fan-in. It follows that min-plus inner products can be computed in $\AC^0$, and the new algorithm manipulates such circuits at the bit-level. (This also means the approach is highly non-black-box, and not subject to lower bounds based on additions and comparisons alone; this is necessary, due to Kerr's $\Omega(n^3)$ lower bound.) $\AC^0$ operations are very structured and have many known limitations~(starting with \cite{Ajtai83,FSS84}). Circuit lower bound techniques often translate algorithmically into nice methods for manipulating circuits.

Razborov~\cite{Razborov} and Smolensky~\cite{Smolensky87} showed how to randomly reduce size-$s$ and depth-$d$ $\AC^0$ circuits with XOR gates to multivariate polynomials over $\F_2$ with $2^{(\log s)^{O(d)}}$ monomials and approximate functionality. We show how elements of their reduction can be applied to randomly translate min-plus inner products of $\ell$-length vectors into $\F_2$ inner products of $n^{0.1}$-length vectors, where $\ell = 2^{(\log n)^{\delta}}$ for some $\delta > 0$. (The straightforward way of applying this reduction also introduces a $\poly(\log M)$ multiplicative factor.) This allows for an efficient reduction from min-plus matrix multiplication of $n \times \ell$ and $\ell \times n$ matrices to a small number of $n \times n^{0.1}$ and $n^{0.1}\times n$ matrix multiplies over $\F_2$. But such \emph{rectangular} matrix multiplications can be computed in $n^2 \cdot \poly(\log n)$ arithmetic operations, using a method of Coppersmith~\cite{Coppersmith82}.\footnote{Technically speaking, Coppersmith only proves a bound on the \emph{rank} of matrix multiplication under these parameters; in prior work~\cite{Williams11} the author described at a high level how Coppersmith's rank bound translates into a full algorithm. An extended exposition of Coppersmith's algorithm is given in Appendix~\ref{coppersmith-appendix}.}
It follows that min-plus matrix multiplication of $n \times \ell$ and $\ell \times n$ matrices is in $n^2 \cdot \poly(\log n)$ time. (There are, of course, many details being glossed over; they will come later.)

This algorithm for rectangular min-plus product can be extended to a product of $n \times n$ matrices in a standard way, by partitioning the matrices into $n/\ell$ products of $n \times \ell$ and $\ell \times n$, computing each product separately, then directly comparing the $n/\ell$ minima found for each of the $n^2$ entries. All in all, we obtain an algorithm for min-plus matrix product running in $\tilde{O}(n^3/\ell+n^2\ell \log M)$ time. Since $\ell = 2^{(\log n)^{\delta}} \gg \log^c n$ for all constants $c$, the $\poly(\log n)$ factors can be absorbed into a bound of $\tilde{O}(n^3/2^{\Omega(\log n)^{\delta}}+n^{2+o(1)})$.\footnote{This $\tilde{O}$ hides not only $\poly(\log n)$ but also $\poly(\log M)$ factors.}

By integrating ideas from prior APSP work, the algorithm for rectangular min-plus product can be improved to a strongly polynomial running time, resulting in Theorem~\ref{main}. First, a standard trick in the literature due to Fredman~\cite{Fredman} permits us to replace the arbitrary entries in the matrices with $O(\log n)$-bit numbers after only $n^{2+o(1)}$ additions and comparisons of the (real-valued) entries; this trick also helps us avoid translating the additions into $\AC^0$. Then we construct a low-depth AND/XOR/NOT circuit for computing the minima of the quantities produced by Fredman's trick, using Razborov-Smolensky style arguments to probabilistically translate the circuit into a multivariate polynomial over $\F_2$ which computes it, with high probability. With care, the polynomial can be built to have relatively few monomials, leading to a better bound.

\subsection{Applications}

The running time of the new APSP algorithm can be extended to many other problems; here we illustrate a few. For notational simplicity, let $\ell(n) = \Theta((\log n)^{1/2})$  be such that APSP is in $n^3/2^{\ell(n)}$ time, according to Theorem~\ref{main}.  It follows from the reductions of~\cite{VW10} (and folklore) that:

\begin{corollary} The following problems are all solvable in $n^3/2^{\Omega(\ell(n))}$ time on the real RAM.
\begin{compactitem}
\item \emph{Metricity}: Determine whether an $n \times n$ matrix over $\R$ defines a metric space on $n$ points.
\item \emph{Minimum weight triangle}: Given an $n$-node graph with real edge weights, compute $u,v,w$ such that $(u,v),(v,w),(w,u)$ are edges and the sum of edge weights is minimized.
\item \emph{Minimum cycle}: Given an $n$-node graph with real positive edge weights, find a cycle of minimum total edge weight.
\item \emph{Second shortest paths}: Given an $n$-node directed graph with real positive edge weights and two nodes $s$ and $t$, determine the second shortest simple path from $s$ to $t$.
\item \emph{Replacement paths}: Given an $n$-node directed graph with real positive edge weights and a shortest path $P$ from node $s$ to node $t$, determine for each edge $e \in P$ the shortest path from $s$ to $t$ in the graph with $e$ removed.
\end{compactitem}
\end{corollary}

Faster algorithms for some \emph{sparse} graph problems also follow from Theorem~\ref{main}. An example is that of finding a minimum weight triangle in a sparse graph:

\begin{theorem}\label{sparsemintriangle} For any $m$-edge weighted graph, a minimum weight triangle can be found in $m^{3/2}/2^{\Omega(\ell(m))}$ time.
\end{theorem}

Bremner et al.~\cite{Necklaces} show that faster algorithms for $(\min,+)$ matrix product imply faster algorithms for computing
 the $(\min,+)$ convolution of two vectors $x, y \in (\Z \cup \{-\infty\})^n$, which is the vector in $(\Z \cup \{-\infty\})^n$
 defined as \[(x \odot y)[i] = \min_{k=1}^i (x[k]+y[i-k]).\] In other words, this is the usual discrete convolution of two vectors in $(\min,+)$ algebra.

\begin{corollary}[\cite{Necklaces}] The $(\min,+)$ convolution of a length-$n$ array is in $n^2/2^{\Omega(\ell(n))}$ time.
\end{corollary}

Is it possible that the approach of this paper can be extended to give a ``truly subcubic'' APSP algorithm, running in $n^{3-\eps}$ time for some $\eps > 0$? If so, we might require an even more efficient way of representing min-plus inner products as inner products over the integers. Very recently, the author discovered a way to efficiently evaluate large depth-two linear threshold circuits~\cite{Williams13thr} on many inputs. The method is general enough that, if min-plus inner product can be efficiently implemented with depth-two threshold circuits, then truly subcubic APSP follows. For instance:

\begin{theorem}\label{depth2APSP} Let $M > 1$ be an integer. Suppose the $(\min,+)$ inner product of two $n$-vectors with entries in $(\Z \cap [0,M])\cup\{\infty\}$ has polynomial-size depth-two threshold circuits with weights of absolute value at most $2^{\poly(\log M)}\cdot 2^{n^2}$, constructible in polynomial time. Then for some $\eps > 0$, APSP is solvable on the word RAM in $n^{3-\eps}\cdot \poly(\log M)$ time for  edge weights in $\Z \cap [0,M]$.
\end{theorem}

To phrase it another way, the hypothesis that APSP is \emph{not} in truly subcubic time implies interesting circuit lower bounds.

\paragraph{Outline of the rest.} In Section~\ref{quick}, we try to provide a relatively succinct exposition of how to solve APSP in less than $n^3/\log^k n$ time for all $k$, in the case where the edge weights are not too large (e.g., at most $\poly(n)$). In Section~\ref{mainresult} we prove Theorem~\ref{main} in full, by expanding considerably on the arguments in Section~\ref{quick}. In Section~\ref{apps} we illustrate one of the many applications, and consider the possibility of extending our approach to a truly subcubic algorithm for APSP. We conclude in Section~\ref{conclusion}.

\section{A relatively short argument for faster APSP}
\label{quick}

We begin with a succinct exposition of a good algorithm for all-pairs shortest paths, at least in the case of reasonable-sized weights. This will illustrate most of the main ideas in the full algorithm.

\begin{theorem}\label{simpleAPSP} There is a deterministic algorithm for APSP which, for some $\delta > 0$, runs in time \[\frac{n^3 \cdot \log M \cdot \poly(\log \log M)}{2^{\Omega(\log n)^{\delta}}}\] on $n$-node graphs with edge weights from $([0,M] \cap \Z)\cup\{\infty\}$.
\end{theorem}

To simplify the presentation, we will not be explicit in our choice of $\delta$ here; that lovely torture will be postponed to the next section. Another mildly undesirable property of Theorem~\ref{simpleAPSP} is that the bound is only meaningful for $M \leq 2^{2^{\eps(\log n)^{\delta}}}$ for sufficiently small $\eps > 0$. So this is not the most general bound one could hope for, but it is effective when the edge weights are in the range $\{0,1,\ldots,\poly(n)\}$, which is already a difficult case for present algorithms. The $(\log M)^{1+o(1)}$ factor will be eliminated in the next section.

Let's start by showing how $\AC^0$ circuit complexity is relevant to the problem. Define ${\cal W} := ([0,M] \cap \Z)\cup \{\infty\}$; intuitively, ${\cal W}$ represents the space of possible weights. Define the min-plus inner product of vectors $u,v \in {\cal W}^d$ to be \[(u\star v) = \min_i (u[i]+v[i]).\] A min-plus matrix multiplication simply represents a collection of all-pairs min-plus inner products over a set of vectors.

\begin{lemma} \label{AC0minplus} Given $u,v \in {\cal W}^d$ encoded as $O(d \log M)$-bit strings, $(u \star v)$ is computable with constant-depth AND/OR/NOT circuits of size $(d \log M)^{O(1)}$. That is, the min-plus inner product function is computable in $\AC^0$ for every  $d$ and $M$.
\end{lemma}

The proof is relatively straightforward; it is given in Appendix~\ref{AC0minplus-appendix} for completeness. Next, we show that a small $\AC^0$ circuit can be quickly evaluated on all pairs of inputs of one's choice. The first step is to deterministically reduce $\AC^0$ circuits to depth-two circuits with a symmetric function (a multivariate Boolean function whose value only depends on the sum of the variables)  computing the output gate and AND gates of inputs (or their negations) on the second layer. Such circuits are typically called $\SYM^+$ circuits~\cite{BT94}. It is known that constant-depth circuits with AND, OR, NOT, and MOD$m$ gates of size $s$ (a.k.a. $\ACC$ circuits) can be efficiently translated into $\SYM^+$ circuits of size $2^{(\log s)^c}$ for some constant $c$ depending on the depth of the circuit and the modulus $m$:\footnote{A MOD$m$ gate outputs $1$ if and only if the sum of its input bits is divisible by $m$.}

\begin{lemma}[\cite{Yao90,BT94,Allender-Gore94}] \label{Beigel-Tarui}
There is an algorithm $A$ and $f : \N\times \N \rightarrow \N$ such that given any size-$s$ depth-$e$ circuit $C$ with AND, OR, and $\text{MOD}m$ gates of unbounded fan-in, $A$ on $C$ runs in $2^{O(\log^{f(e,m)} s)}$ time and outputs an equivalent $\SYM^+$ circuit of $2^{O(\log^{f(e,m)} s)}$ gates.

Moreover, given the number of ANDs in the circuit evaluating to $1$, the symmetric function itself can be evaluated in $(\log s)^{O(f(e,m))}$ time.
\end{lemma}

It is easy to see that this translation is really converting circuits into multivariate \emph{polynomials} over $\{0,1\}$: the AND gates represent monomials with coefficients equal to $1$, the sum of these AND gates is a polynomial with $2^{O(\log^{f(e,m)} s)}$ monomials, and the symmetric function represents some efficiently computable function from $\Z$ to $\{0,1\}$.

The second step is to quickly evaluate these polynomials on many chosen inputs, using rectangular matrix multiplication. Specifically, we require the following:

\begin{lemma}[Coppersmith~\cite{Coppersmith82}]\label{rectangular} For all sufficiently large $N$, multiplication of an $N \times N^{.172}$ matrix with an $N^{.172} \times N$ matrix can be done in $O(N^2 \log^2 N)$ arithmetic operations.\footnote{See Appendix~\ref{coppersmith-appendix} for a detailed exposition of this algorithm.}\end{lemma}

\begin{theorem}\label{polyeval} Let $p$ be a $2k$-variate polynomial over the integers (in its monomial representation) with $m \leq n^{0.1}$ monomials, along with $A, B \subseteq \{0,1\}^{k}$ such that $|A|=|B|=n$. The polynomial $p(a_1,\ldots,a_k,b_1,\ldots,b_k)$ can be evaluated over all points $(a_1,\ldots,a_k,b_1,\ldots,b_k) \in A \times B$ in $n^2 \cdot \poly(\log n)$ arithmetic operations.
\end{theorem}

Note that the obvious polynomial evaluation algorithm would require $n^{2.1}$ arithmetic operations.

\begin{proof} Think of the polynomial $p$ as being over two sets of variables, $X=\{x_1,\ldots,x_k\}$ and $Y = \{y_1,\ldots,y_k\}$. First, we construct two matrices $M_1 \in \Z^{n \times m}$ and $M_2 \in \Z^{m \times n}$ as follows. The rows $i$ of $M_1$ are indexed by the elements $r_1,\ldots,r_{|A|}\in \{0,1\}^k$ of $A$, and the columns $j$ are indexed by the monomials $p_1,\ldots,p_m$ of $p$. Let $p_i|_{X}$ denote the monomial $p_i$ restricted to the variables $x_1,\ldots,x_k$ (including the coefficient of $p_i$), and $p_i|_{Y}$ denote the product of all variables from $y_1,\ldots,y_k$ appearing in $p_i$ (here the coefficient of $p_i$ is \emph{not} included). Observe that $p_i|_{X}\cdot p_i|_{Y}=p_i$. Define
$M_1[i,j] := p_i|_{X}(r_j)$. The rows of $M_2$ are indexed by the monomials of $p$, the columns are indexed by the elements $s_1,\ldots,s_{|B|}\in\{0,1\}^k$ of $B$, and 
$M_2[i,j] := p_j|_{Y}(s_i)$. Observe that $(M_1 \cdot M_2)[i,j] = p(r_i,s_j)$. Applying Lemma~\ref{rectangular} for $n \times n^{0.1}$ and $n^{0.1} \times n$ matrices, $M_1 \cdot M_2$ is computable in $n^2 \cdot \poly(\log n)$ operations.
\end{proof}

Putting the pieces together, we obtain our ``warm-up'' APSP algorithm:

\begin{proofof}{Theorem~\ref{simpleAPSP}} Let $A$ and $B$ be $n \times n$ matrices over ${\cal W} = ([0,M] \cap \Z)\cup\{\infty\}$. We will show there is a universal $c \geq 1$ such that we can min-plus multiply an arbitrary $n \times d$ matrix $A'$ with an arbitrary $d \times n$ matrix $B'$ in $n^2 \cdot \poly(\log n, \log \log M)$ time, for $d \leq \frac{2^{(0.1 \log n)^{1/c}}}{\log M}$.\footnote{Note that, if $M \geq 2^{2^{(0.1 \log n)^{1/c}}}$, the desired running time is trivial to provide.} By decomposing the matrix $A$ into a block row of $n/d$ $n\times d$ matrices, and the matrix $B$ into a block column of $n/d$ $d \times n$ matrices, it follows that we can min-plus multiply $n \times n$ and $n \times n$ matrices in time \[(n^3 \cdot \log M\cdot  \poly(\log \log M))/2^{\Omega(\log n)^{1/c}}.\]

So let $A'$ and $B'$ be $n \times d$ and $d \times n$, respectively. Each row of $A'$ and column of $B'$ defines a min-plus inner product of two $d$-vectors $u,v \in {\cal W}^d$. By Lemma~\ref{AC0minplus}, there is an $\AC^0$ circuit $C$ of size $(d \log M)^{O(1)}$ computing $(u \star v)$ for all such vectors $u,v$. By Theorem~\ref{Beigel-Tarui}, that circuit $C$ can be simulated by a polynomial $p : \{0,1\}^{O(d \log M)} \rightarrow \Z$ of at most $K=2^{(\log(d \log M))^{c}}$ monomials for some integer $c \geq 1$, 
followed by the efficient evaluation of a function from $\Z$ to $\{0,1\}$ on the result.
For $K \leq n^{0.1}$, Theorem~\ref{polyeval} applies, and we can therefore compute all pairs of min-plus inner products consisting of rows $A'$ and columns of $B'$ in time $n^2 \cdot \poly(\log n)$ operations over $\Z$, obtaining their min-plus matrix product.

But $K \leq n^{0.1}$ precisely when $(\log(d \log M))^c \leq 0.1 \log n$, i.e., \[d \leq \frac{2^{(0.1 \log n)^{1/c}}}{\log M}.\] Therefore, we can compute an $n \times \frac{2^{(0.1 \log n)^{1/c}}}{\log M}$ and $\frac{2^{(0.1 \log n)^{1/c}}}{\log M} \times n$ min-plus matrix product in $n^2 \cdot \poly(\log n)$ arithmetic operations over $\Z$. To ensure the final time bound, observe that each coefficient of the polynomial $p$ has bit complexity at most $(\log(d \log M))^c \leq (\log n + \log \log M)^c \leq \poly(\log n,\log \log M)$ (there could be multiple copies of the same AND gate in the $\SYM^+$ circuit), hence the integer output by $p$ has at most $\poly(\log n,\log \log M)$ bit complexity as well. Evaluating the symmetric function on each entry takes $\poly(\log n,\log \log M)$ time. Hence the aforementioned rectangular min-plus product is in $n^2 \cdot \poly(\log n, \log \log M)$ time, as desired.
\end{proofof}

\section{Proof of The Main Theorem}\label{mainresult}

In this section, we establish Theorem~\ref{main}. This algorithm will follow the basic outline of Section~\ref{quick}, but we desire a strongly polynomial time bound with a reasonable denominator. To achieve these goals, we incorporate Fredman's trick into the argument, and we carefully apply the polynomials of Razborov and Smolensky for $\AC^0$ circuits with XOR gates. Here, the final polynomials will be over the field $\F_2 = \{0,1\}$ instead of $\Z$.

Let $A$ be an $n \times d$ matrix with entries from ${\cal W} := ([0,M] \cap \Z) \cup \{\infty\}$, and let $B$ be an $d \times n$ matrix with entries from ${\cal W}$. We wish to compute \[C[i,j] = \min_{k=1}^d (A[i,k]+B[k,j]).\] First, we can assume without loss of generality that for all $i,j$, there is a unique $k$ achieving the minimum $A[i,k]+B[k,j]$. One way to enforce this is to change all initial $A[i,j]$ entries at the beginning to $A[i,j]\cdot (n+1) + j$, and all $B[i,j]$ entries to $B[i,j]\cdot(n+1)$, prior to sorting. These changes can be made with only $O(\log n)$ additions per entry; e.g., by adding $A[i,j]$ to itself for $O(\log n)$ times. Then, $\min_k A[i,k]+B[k,j]$ becomes \[\min_k (A[i,k]+B[k,j])\cdot (n+1) + k^{\star},\] where $k^{\star}$ is the minimum integer achieving $\min_k A[i,k]+B[k,j]$. 

Next, we encode a trick of Fredman~\cite{Fredman} in the computation; his trick is simply that \[A[i,k]-A[i,k'] \leq B[k',j]-B[k,j] \text{~if and only if~} A[i,k]+B[k,j] \leq A[i,k']+B[k',j].\] This subtle trick has been applied in most prior work on faster APSP. It allows us to ``prepare'' $A$ and $B$ by taking many differences of entries, before making explicit comparisons between entries. Namely, we construct matrices $A'$ and $B'$ which are $n \times d^2$ and $d^2 \times n$. The columns of $A'$ and rows of $B'$ are indexed by pairs $(k,k')$ from $[d]^2$. We define: \[A'[i,(k,k')] := A[i,k]-A[i,k'] \text{~and~} B'[(k,k'),j] := B[k',j]-B[k,j].\] Observe that $A'[i,(k,k')] \leq B'[(k,k'),j]$ if and only if $A[i,k]+B[k,j] \leq A[i,k']+B[k',j]$.

For each column $(k,k')$ of $A'$ and corresponding row $(k,k')$ of $B'$, sort the $2n$ numbers in the set \[S_{(k,k')} = \{A'[i,(k,k')], B'[(k,k'),i] ~|~ i=1,\ldots,n\},\] and replace each $A'[i,(k,k')]$ and $B[(k,k'),j]$ by their rank in the sorted order on $S_{(k,k')}$, breaking ties arbitrarily (giving $A$ entries precedence over $B$ entries). Call these new matrices $A''$ and $B''$. The key properties of this replacement are:\begin{enumerate}
\item All entries of $A''$ and $B''$ are from the set $\{1,\ldots,2n\}$.
\item $A''[i,(k,k')] \leq B''[(k,k'),j]$ if and only if $A'[i,(k,k')] \leq B'[(k,k'),j]$. That is, the outcomes of all comparisons have been preserved.
\item For every $i,j$, there is a unique $k$ such that $A''[i,(k,k')] \leq B''[(k,k'),j]$ for all $k'$; this follows from the fact that there is a unique $k$ achieving the minimum $A[i,k]+B[k,j]$. 
\end{enumerate}
This replacement takes $\tilde{O}(n\cdot d^2\cdot \log M)$ time on a word RAM, and $O(n\cdot d^2 \cdot \log n)$ on the real RAM.\footnote{As observed by Zwick~\cite{Zwick04}, we do not need to allow for unit cost subtractions in the model; when we wish to compare two quantities $x-y$ and $a-b$ in the above, we simulate this by comparing $x+b$ and $a+y$, as in Fredman's trick.}

To determine the $(\min,+)$ product of $A$ and $B$, by the proof of Lemma~\ref{AC0minplus} (in the appendix) it suffices to compute for each $i,j=1,\ldots,n$, and $\ell = 1,\ldots,\log d$, the logical expression

\[P(i,j,\ell) =
\bigvee_{\substack{k =1,\ldots,d \\\text{$\ell$th bit of $k$ is $1$}}} \bigwedge_{~k' \in \{1,\ldots,d\}} \text{\bf [}A''[i,(k,k')] \leq B''[(k,k'),j]\text{\bf ]}.\] Here we are using the notation that, for a logical expression $Q$, the expression {\bf [}$Q${\bf ]} is either $0$ or $1$, and it is $1$ if and only if $Q$ is true.

We claim that $P(i,j,\ell)$ equals the $\ell$th bit of the smallest $k^{\star}$ such that $\min_{k} A[i,k]+B[k,j] = A[i,k^{\star}]+B[k^{\star},j]$. In particular, by construction of $A''$, the $\wedge$ in the expression $P(i,j,\ell)$ is true for a given $k^{\star}$ if and only if for all $k'$ we have $A[i,k^{\star}] + B[k^{\star},j] \leq A[i,k'] + B[k',j]$, which is true if and only if $\min_{k''=1}^d A[i,k''] + B[k'',j] = A[i,k^{\star}] + B[k^{\star},j]$ and $k$ is the smallest such integer (the latter being true due to our sorting constraints). Finally, $P(i,j,\ell)$ is $1$ if and only if the $\ell$th bit of this particular $k^{\star}$ is $1$. This proves the claim.

We want to translate $P(i,j,\ell)$ into an expression we can efficiently evaluate arithmetically. We will do several manipulations of $P(i,j,\ell)$ to yield polynomials over $\F_2$ with a ``short'' number of monomials. Observe that, since there is always exactly one such $k^{\star}$ for every $i,j$, exactly \emph{one} of the $\wedge$ expressions in $P(i,j,\ell)$ is true for each fixed $i,j,\ell$. Therefore we can replace the $\vee$ in $P(i,j,\ell)$ with an XOR (also denoted by $\oplus$):

\[P(i,j,\ell) =
\bigoplus_{\substack{k =1,\ldots,d \\\text{$\ell$th bit of $k$ is $1$}}} \bigwedge_{~k' \in \{1,\ldots,d\}} \text{\bf [}A''[i,(k,k')] \leq B''[(k,k'),j]\text{\bf ]}.\]

This is useful because XORs are ``cheap'' in an $\F_2$ polynomial, whereas ORs can be expensive. Indeed, an XOR is simply addition over $\F_2$, while AND (or OR) involves multiplication which can lead to many monomials.

In the expression $P$, there are $d$ different ANDs over $d$ comparisons. In order to get a ``short'' polynomial, we need to reduce the fan-in of the ANDs. Razborov and Smolensky proposed the following construction: for an AND over $d$ variables $y_1,\ldots,y_d$, let $e \geq 1$ be an integer, choose independently and uniformly at random $e\cdot d$ bits $r_{1,1},\ldots,r_{1,d},r_{2,1},\ldots,r_{2,d}, ~\ldots~, r_{e,1},\ldots,r_{e,d} \in \{0,1\}$, and consider the expression \[E(y_1,...,y_d) = \bigwedge_{i=1}^e \left(1+\bigoplus_{j=1}^d r_{i,j}\cdot(y_j + 1)\right),\] where $+$ corresponds to addition modulo $2$. Note that when the $r_{i,j}$ are fixed constants, $E$ is an AND of $e$ XORs of at most $d+1$ variables $y_j$ along with possibly the constant $1$.
\begin{claim}[Razborov~\cite{Razborov}, Smolensky~\cite{Smolensky87}]\label{razsmo}
For every fixed $(y_1,...,y_d)\in\{0,1\}^d$, \[\Pr_{r_{i,j}}[ E(y_1,...,y_d) = y_1 \wedge \cdots \wedge y_d] \geq 1-1/2^e.\]
\end{claim}

For completeness, we give the simple proof. For a given point $(y_1,\ldots,y_d)$, first consider the expression $F_i = 1+\oplus_{j=1}^d r_{i,j}\cdot (y_j+1)$. If $y_1 \wedge \cdots \wedge y_d = 1$, then $(y_j+1)$ is $0$ modulo $2$ for all $j$, and hence $F_i = 1$ with probability $1$. If $y_1 \wedge \cdots \wedge y_d = 0$, then there is a subset $S$ of $y_j$'s which are $0$, and hence a subset $S$ of $(y_j+1)$'s that are $1$. The probability we choose $r_{i,j}=1$ for an odd number of the $y_j$'s in $S$ is at exactly $1/2$. Hence the probability that $F_i = 0$ in this case is exactly $1/2$.

Since $E(y_1,\ldots,y_d) = \wedge_{i=1}^e F_i$, it follows that if $y_1 \wedge \cdots \wedge y_d = 1$, then $E = 1$ with probability $1$. Since the $r_{i,j}$ are independent, if $y_1 \wedge \cdots \wedge y_d = 0$, then the probability is only $1/2^e$ that for all $i$ we have $r_{i,j}=1$ for an odd number of $y_j=0$. Hence the probability is $1-1/2^e$ that some $F_i(y_1,\ldots,y_d)=0$, completing the proof.

Now set $e = 2+\log d$, so that $E$ fails on a point $y$ with probability at most $1/(4d)$. Suppose we replace each of the $d$ ANDs in expression $P$ by the expression $E$, yielding:
\[P'(i,j,\ell) =
\bigoplus_{\substack{k =1,\ldots,d \\\text{$\ell$th bit of $k$ is $1$}}} E(\text{\bf [}A''[i,(k,1)] \leq B''[(k,1),j]\text{\bf ]},\ldots,\text{\bf [}A''[i,(k,k')] \leq B''[(k,k'),j]\text{\bf ]}).\]
By the union bound, the probability that the (randomly generated) expression $P'$ differs from $P$ on a given row $A''[i,:]$ and column $B''[:,j]$ is at most $1/4$.

Next, we open up the $d^2$ comparisons in $P$ and simulate them with low-depth circuits. Think of the entries of $A''[i,(k,k')]$ and $B''[(k,k'),j]$ as bit strings, each of length $t = 1+\log n$. To check whether $a \leq b$ for two $t$-bit strings $a = a_1,...,a_t$ and $b = b_1,...,b_t$ construed as positive integers in $\{1,\ldots,2^t\}$, we can compute (from Lemma~\ref{AC0minplus})\begin{eqnarray*}
LEQ(a,b) &=& \left(\bigwedge_{i=1}^{t} (1 + a_i + b_i)\right)\\ 
 & & \oplus \bigoplus_{i=1}^{t} \left((1 + a_i) \wedge b_i \wedge \bigwedge_{j=1}^{i-1} (1 + a_j + b_j)\right)\end{eqnarray*} 
 where $+$ again stands for addition modulo $2$. (We can replace the outer $\vee$ with a $\oplus$, because at most one of the $t$ expressions inside of the $\oplus$ can be true for any $a$ and $b$.)

The $LEQ$ circuit is an XOR of $t+1$ ANDs of fan-in $\leq t$ of XORs of fan-in at most 3. Applying Claim~\ref{razsmo}, we replace the ANDs with a randomly chosen expression $E'(e_1,\ldots,e_{t})$, which is an AND of fan-in $e'$ (for some parameter $e'$ to be determined) of XORs of $\leq t$ fan-in. The new expression $LEQ'$ now has the form \begin{equation}\label{formLEQ}\bigoplus_{t+1}\left[\bigwedge_{e'}\left[\bigoplus_{\leq t}\left[\text{2 $\oplus$ gates}\right]\right]\right];\end{equation}
that is, we have an XOR of $t+1$ fan-in, of ANDs of fan-in $e'$, of XORs of $\leq t$ fan-in, of XORs of fan-in at most 3. 

In fact, an anonymous STOC referee pointed out that, by performing additional preprocessing on the matrices $A''$ and $B''$, we can reduce the $LEQ'$ expression further, to have the form \[\bigoplus_{t+1}\left[\bigwedge_{e'}\left[\text{2 $\oplus$ gates}\right]\right].\] This reduction will be significant enough to yield a better denominator in the running time. (An earlier version of the paper, without the following preprocessing, reported a denominator of $2^{\Omega(\log n / \log \log n)^{1/2}}$.) Each term of the form ``$\bigoplus_{\leq t}\left[\text{2 $\oplus$ gates}\right]$'' in \eqref{formLEQ} can be viewed an XOR of three quantities: an XOR of a subset of $O(\log n)$ variables $a_i$ (from the matrix $A''$), another XOR of a subset of $O(\log n)$ variables $b_j$ (from the matrix $B''$), and a constant (0 or 1). Given the random choices to construct the expression $E'$, we first compute the $(t+1)e'$ XORs over just the entries from the matrix $A''$ in advance, for all $nd^2$ entries in $A''$, and separately compute the set of $(t+1)e'$ XORs for the $nd^2$ entries in $B$, in $\tilde{O}(n d^2 \cdot (t+1)e')$ time. Once precomputed, these XOR values will become the values of variables in our polynomial evaluation later. For each such XOR over an appropriate subset $S$ of the $a_j$'s (respectively, some subset $T$ of the $b_j$'s), we introduce new variables $a'_S$ (and $b'_T$), and from now on we think of evaluating the equivalent polynomial over these new $a'_S$ and $b'_T$ variables, which has the form \[\bigoplus_{t+1}\left[\bigwedge_{e'}\left[\text{2 $\oplus$ gates}\right]\right].\]

Combining the two consecutive layers of XOR into one, and applying the distributive law over $\F_2$ to the AND, $LEQ'$ is equivalent to a degree-$e'$ polynomial $Q$ over $\F_2$  with at most $m = (t+1)\cdot 3^{e'}$ monomials (an XOR of fan-in at most $m$ of ANDs of fan-in at most $e'$). By the union bound, since the original circuit for $LEQ(a,b)$ contains only $t+1$ AND gates, and the probability of error of $E'$ is at most $1/2^{e'}$, we have that for a fixed pair of strings $(a,b)$, $LEQ(a,b) = LEQ'(a,b)$ with probability at least $1-(t+1)/2^{e'}$.

Recall in the expression $P'$, there are $d^2$ comparisons, and hence $d^2$ copies of the $LEQ$ circuit are needed. Setting \[e' = 3+2\log d + \log t,\] we ensure that, for a given row $i$, column $j$, and $t$ for $P'$, $d^2$ copies of the $LEQ'$ circuit give the same output as $LEQ$ with probability at least $3/4$.

Hence we have a polynomial $Q$ in at most $m' = (t+1)\cdot 3^{3+2\log d + \log t}$ monomials, each of degree at most $2t$, that can accurately computes all comparisons in $P'$ on a given point, with probability at least $3/4$. Plugging $Q$ into the circuit for $P'$, the expression $P''(i,j,\ell)$ now has the form: 
\[\begin{array}{l}
\text{An XOR of $\leq d$ fan-in,}\\
~~~~~ \text{ANDs of $1+\log d$ fan-in,}\\
~~~~~ ~~~~~ \text{XORs of $\leq d+1$ fan-in,}\\
~~~~~ ~~~~~ ~~~~~~ \text{XORs of $\leq m'$ fan-in,}\\
~~~~~ ~~~~~ ~~~~~~ ~~~~~~ \text{ANDs of $e'$ variables.}
\end{array}\] (The second and third layers are the $E$ circuits; the fourth and fifth layers are the polynomial $Q$ applied to various rows and columns.) Merging the two consecutive layers of XORs into one XOR of fan-in $\leq (d+1)m'$, and applying distributivity to the ANDs of $\leq 1+\log d$ fan-in, we obtain a polynomial $Q'_{i,j,\ell}$ over $\F_2$ with a number of monomials at most \begin{eqnarray*}\lefteqn{d\cdot((d+1)m')^{1+\log d}}\\
& \leq& d\cdot((d+1)\cdot (t+1)\cdot 3^{3+2\log d + \log t})^{1+\log d}.\end{eqnarray*} Further simplifying, this quantity is at most \begin{equation}\label{monomials-final} 2^{(1+\log d)\cdot(\log(d+1) + \log (t+1) + (\log 3)(3 + 2\log d + \log t))}.\end{equation}

Let $m''$ denote the quantity in \eqref{monomials-final}. Provided $m'' \leq n^{0.1}$, we will be able to apply a rectangular matrix multiplication in the final step. This is equivalent to
\begin{eqnarray}\label{logmonomials}
\log_2(m'') \leq 0.1 \log n. \end{eqnarray}
Recall $t = 1+\log n$, and note that $\log_2(m'')$ expands to a sum of various powers of logs. For $d \geq t$, the dominant term in $\log_2(m'')$  is $2(\log^2 d)(\log 3) \leq O((\log d)^2)$. Choosing \[d = 2^{\delta\cdot(\log n)^{1/2}}\] for sufficiently small $\delta > 0$, inequality \eqref{logmonomials} will be satisfied, and the number $m''$ will be less than $n^{0.1}$.

Finally, we apply Coppersmith's rectangular matrix multiplication (Lemma~\ref{rectangular}) to evaluate the polynomial $Q'_{i,j,\ell}$ on all $n^2$ pairs $(i,j)$ in $n^2\cdot \poly(\log n)$ time. For a fixed $\ell=1,\ldots,\log d$, the outcome is a matrix product $D_{\ell}$ such that, for every $(i,j)\in[n]^2$ and for each $\ell = 1,\dots,\log d$, \begin{eqnarray*}\Pr[D_{\ell}[i,j] = P(i,j,\ell)] &=& \Pr[D_{\ell}[i,j]\text{ is the $\ell$th bit of the smallest $k^{\star}$ such that}\\
& & \text{~~~~~~~~~~~~~~~~ $A[i,k^{\star}]+B[k^{\star},j] = \min_k(A[i,k]+B[k,j])$]}
\\ &\geq& 3/4.\end{eqnarray*}

Correct entries for all $i,j$ can be obtained with high probability, using a standard ``majority amplification'' trick. Let $c$ be an integer parameter to set later. For every $\ell=1,\ldots,\log d$, choose $c\log n$ independent random polynomials $Q'_{i,j,\ell}$ according to the above process, and evaluate each one on all $i,j\in[n]^2$ using a rectangular matrix product, producing 0-1 matrices $D_{\ell,1},\ldots, D_{\ell,c\log n}$ each of dimension $n \times n$. Let 
\[C_{\ell}[i,j] = MAJ(D_{\ell,1}[i,j],\ldots,D_{\ell,c \log n})[i,j],\] i.e., $C_{\ell}[i,j]$ equals the majority bit of $D_{\ell,1}[i,j],\ldots,D_{\ell,c \log n}[i,j]$.

We claim that $C_{\ell}[i,j]$ equals the desired output for all $i,j,\ell$, with high probability. For every $(i,j) \in [n]^2$, $\ell \in [\log d]$, and $k=1,\ldots,c\log n$, we have $\Pr[D_{\ell,k}[i,j]=P(i,j,\ell)] \geq 3/4$. Therefore for the random variable $X := \sum_{k=1}^{c\log n}$ {\bf [}$D_{\ell,k}[i,j]=P(i,j,\ell)${\bf ]}, we have $E[X] \geq (3c\log n)/4$. In order for the event $MAJ(D_{\ell,1}[i,j],\ldots,D_{\ell,c\log n}) \neq P(i,j,\ell)$ to happen, we must have that $X < (c\log n)/2$.

Recall that if we have independent random variables $Y_i$ that are $0$-$1$ valued with $0 < E[Y_i] < 1$, the random variable $Y := \sum_{i=1}^k Y_i$ satisfies the tail bound \[\Pr\left[Y < (1-\eps)E[Y]\right] \leq e^{-\eps^2 E[Y]/2}\] (e.g., in Motwani and Raghavan~\cite{MR}, this is Theorem 4.2). Applying this bound,
\begin{eqnarray*}
\Pr[C_{\ell}(i,j) \neq P(i,j,\ell)] &=& \Pr[MAJ(D_{\ell,1}[i,j],\ldots,D_{\ell, c\log n}[i,j]) \neq P(i,j,\ell)]\\
& \leq & \Pr\left[X < (c\log n)/2\right]\leq \Pr\left[X < (1-1/3)E[X]\right]\\
& \leq & e^{-(2/3)^2 E[X]/2}=e^{-4E[X]/18}.\end{eqnarray*}

Set $c=18$. By a union bound over all pairs $(i,j)\in[n]^2$ and $\ell\in[\log d]$,\begin{eqnarray*}
\Pr[\text{There are $i,j,\ell$, } C_{\ell} \neq P(i,j,\ell)]
&\leq& (n^2\log d)\cdot e^{-4\log n} \leq (\log d)/n^2.
\end{eqnarray*}

Set $c=18$. By a union bound over all pairs $(i,j)\in[n]^2$ and $\ell\in[\log d]$, \[\Pr[\text{there are $i,j,\ell$, } C_{\ell} \neq P(i,j,\ell)]\leq (n^2\log d)\cdot e^{-4\log n} \leq (\log d)/n^2.\]

Therefore for $d = 2^{\delta(\log n)^{1/2}}$, the algorithm outputs the min-plus product of an $n \times d$ and $d \times n$ matrix in $n^2\cdot \poly(\log n) + n\cdot d^2\cdot(\log M)$ time, with probability at least $1-(\log n)/n^2$.

Applying this algorithm to $n/d$ different $n \times d$ and $d \times n$ min-plus products, the min-plus product of two $n \times n$ matrices is computable in time $n^3/2^{\Omega(\log n)^{1/2}}$ on the real RAM with probability at least $1-(\log n)/n$, by the union bound. (On the word RAM, there is an extra additive factor of $n^{2+o(1)}\cdot \log M$, for the initial application of Fredman's trick.)

\subsection{Derandomizing the algorithm}

The APSP algorithm can be made deterministic with some loss in the running time, but still asymptotically better than $n^3/(\log n)^k$ for every $k$. See Appendix~\ref{detAPSP-appendix} for the proof.

\begin{theorem} \label{detAPSP} There is a $\delta > 0$ and a deterministic algorithm for APSP running in $n^3/2^{(\log n)^{\delta}}$ time on the real RAM.
\end{theorem}

\section{Some Applications}\label{apps}

All applications referred to the introduction follow straightforwardly from the literature, except for possibly:

\begin{reminder}{Theorem~\ref{sparsemintriangle}} For any $m$-edge weighted graph, a minimum weight triangle can be found in $m^{3/2}/2^{\Omega(\ell(m))}$ time.
\end{reminder}

\begin{proof} We follow the high-degree/low-degree trick of Alon, Yuster, Zwick~\cite{Alon-Yuster-Zwick97}. To find a minimum edge-weight triangle with $m$ edges, let $\Delta \in [1,m]$ be a parameter and consider two possible scenarios:
\begin{compactenum}
\item \emph{The min-weight triangle contains a node of degree at most $\Delta$.} Here, $O(m\cdot \Delta)$ time suffices to search for the triangle: try all possible edges $\{u,v\}$ with $\deg(v) \leq \Delta$, and check if there is a neighbor of $v$ which forms a triangle with $u$, recording the triangle encountered of smallest weight.
\item \emph{The min-weight triangle contains only nodes of degree at least $\Delta$.} Let $N$ be the number of nodes of degree at least $\Delta$; by counting, $N \leq 2m/\Delta$. Searching for a min-weight triangle on these $N$ nodes can be done in $O(N^3/2^{\Omega(\ell(N))})$ time, by reduction to $(\min,+)$ matrix multiplication. In particular, one $(\min,+)$ matrix multiply will efficiently compute the weight of the shortest path of two edges from $u$ to $v$, for every pair of nodes $u,v$. We can obtain the minimum weight of any triangle including the edge $\{u,v\}$ by adding the two-edge shortest path cost from $u$ to $v$ with the weight of $\{u,v\}$. Hence this step takes $O\left(\frac{m^3}{\Delta^3 2^{\Omega(\ell(m/\Delta))}}\right)$ time.
\end{compactenum}

To minimize the overall running time, we want \[m\cdot \Delta \approx m^3/(\Delta^3 2^{\Omega(\ell(m/\Delta))}).\] For $\Delta = m^{1/2}/2^{\ell(m)}$, the runtime is $O(m^{3/2}/2^{\Omega(\ell(m))})$.
\end{proof}

\subsection{Towards Truly Subcubic APSP?}

It seems likely that the basic approach taken in this paper can be extended to discover even faster APSP algorithms. Here we outline one concrete direction to pursue.

A $\SYM\circ\THR$ circuit is a logical circuit of three layers: the \emph{input layer} has $n$ Boolean variables, the \emph{middle layer} contains \emph{linear threshold gates} with inputs from the input layer, and the \emph{output layer} is a single gate taking inputs from the middle layer's outputs and computing a Boolean symmetric function, i.e., the output of the function depends only on the number of true inputs. Every linear threshold gate in the circuit with inputs $y_1,\ldots,y_t$ has its own collection of weights $w_1,\ldots,w_t,w_{t+1} \in \Z$, such that the gate outputs  $1$ if and only if $\sum_{i=1}^t w_i \cdot y_i \geq w_{t+1}$ holds.

It is an open frontier in circuit complexity to exhibit explicit functions which are not computable efficiently with $\SYM\circ\THR$ circuits. As far as we know, it could be that huge complexity classes like ${\sf EXP}^{\sf NP}$ have $\SYM\circ\THR$ circuits with only $\poly(n)$ gates. (Allowing exponential weights is crucial: there are lower bounds for depth-two threshold circuits with small weights~\cite{Hajnal93}.)

\begin{reminder}{Theorem~\ref{depth2APSP}} Let $M > 1$ be an integer. Suppose the $(\min,+)$ inner product of two $n$-vectors with entries in $(\Z \cap [0,M])\cup\{\infty\}$ has polynomial-size $\SYM\circ\THR$ circuits with threshold weights of absolute value at most $2^{\poly(\log M)}\cdot 2^{n^2}$, constructible in polynomial time. Then APSP is solvable on the word RAM in $n^{3-\eps}\cdot \poly(\log M)$ time for some $\eps > 0$ for  edge weights in $\Z \cap [0,M]$.
\end{reminder}

That is,  efficient depth-two circuits for $(\min,+)$ inner product would imply a truly subcubic time algorithm for APSP. The proof applies a recent algorithm of the author:

\begin{theorem}[\cite{Williams13thr}]\label{thrthreval} Given a $\SYM\circ\THR$ circuit $C$ with $2k$ inputs and at most $n^{1/12}$ gates with threshold weights of absolute value at most $W_b$, and given two sets $A,B \subseteq \{0,1\}^k$ where $|A|=|B|=n$, we can evaluate $C$ on all $n^2$ points in $A \times B$ using $n^2 \cdot \poly(\log n) + n^{1+1/12}\cdot \poly(\log n,\log W_b)$ time.\end{theorem}

A similar theorem also holds for depth-two threshold circuits ($\THR \circ \THR$). Note the obvious algorithm for the above evaluation problem would take at least $\Omega(n^{2+1/12})$ time.

\begin{proofof}{Theorem~\ref{depth2APSP}}
Assuming the hypothesis of the theorem, there is some $k$ such that the $(\min,+)$ inner product of two $d$-vectors with entries in $([0,M] \cap \Z)\cup\{\infty\}$ can be computed with a depth-two linear threshold circuit of at most $(d\cdot \log M)^k$ gates. Setting $d = \min\{1,n^{1/(12k)}/(\log M)^k\}$, the number of gates in the circuit is bounded by $n^{1/24}$. (For sufficiently large $M$, $d$ will be $1$, but in this case a time bound of $n^{3-\eps}\cdot \poly(\log M)$ for APSP is trivial.) Letting $A$ be the rows of one $n \times d$ matrix $A'$, and letting $B$ be the columns of another $d \times n$ matrix $B'$, Theorem~\ref{thrthreval} says that we can $(\min,+)$-multiply $A'$ and $B'$ with entries from $([0,M] \cap \Z)\cup\{\infty\}$ in $n^2 \cdot \poly(\log n, \log M)$ time.

To compute the $(\min,+)$-multiplication of two $n\times n$ matrices, we reduce it into $n/d$ multiplies of $n \times d$ and $d \times n$ (as in Theorems~\ref{simpleAPSP} and~\ref{main}), resulting in an algorithm running in time $O(n^{3-1/(12k)}\cdot (\log M)^k)$.

In a graph with edge weights in $\Z \cap [0,M]$, the shortest path between nodes $u$ and $v$ either has length at most $nM$, or it is $\infty$. The above argument shows we can compute min-plus matrix products with entries up to $nM$ in time $n^{3-\eps}\cdot \poly(\log nM) \leq n^{3-\eps'} \poly(\log M)$, for some $\eps, \eps' > 0$. Therefore, APSP can be computed in the desired time, since the necessary min-plus matrix products can be performed in the desired time.
\end{proofof}

\section{Discussion}\label{conclusion}

The method of this paper is generic: the main property of APSP being used is that min-plus inner product and related computations are in $\AC^0$. Other special matrix product operations with ``inner product'' definable in $\AC^0$ (or even $\ACC$) are also computable in $n^3/2^{(\log n)^{\delta}}$ time, as well. (Note that $\AC^0$ by itself is not enough: one must also be able to reduce inner products on vectors of length $n$ to $\tilde{O}(n/d)$ inner products on vectors of length at most $d^{\poly(\log d)}$, as is the case with $(\min,+)$ inner product.) Other fundamental problems have simple algorithms running in time $n^k$ for some $k$, and the best known running time is stuck at  $n^k/\log^c n$ for some $c \leq 3$. (The phrase ``shaving logs'' is often associated with this work.) It would be very interesting to find other basic problems permitting a ``clean shave'' of all polylog factors from the runtime. Here are a few specific future directions.

{\bf 1. Subquadratic 3SUM.} Along with APSP, the 3SUM problem is another notorious polynomial-time solvable problem: \emph{given a list of integers, are there three which sum to zero}? For lists of $n$ numbers, an $O(n^2)$ time algorithm is well-known, and the conjecture that no $n^{1.999}$ time algorithm exists is  significant in computational geometry and data structures, with many intriguing consequences~\cite{GO95,BH01,SEO03,Patrascu10,VW13}. Baran, Demaine, and Patrascu~\cite{BDP08} showed that 3SUM is in about $n^2/\log^2 n$ time (omitting $\poly(\log \log n)$ factors). Can this be extended to  $n^2/2^{(\log n)^{\delta}}$ time for some $\delta > 0$?  It is natural to start with solving Convolution-3SUM, defined by Patrascu~\cite{Patrascu10} as: \emph{given an array $A$ of $n$ integers, are there $i$ and $j$ such that $A[i]+A[j]=A[i+j\pmod n]$?} Although this problem looks superficially easier than 3SUM, Patrascu showed that if Convolution-3SUM is in $n^2/(f(n\cdot f(n)))^2$ time then 3SUM is in $n^2/f(n)$ time. That is, minor improvements for Convolution-3SUM would yield similar improvements for 3SUM.

{\bf 2. Subquadratic String Matching.} There are many problems involving string matching and alignment which are solvable using dynamic programming in $O(n^2/\log n)$ time, on strings of length $n$. A prominent example is computing the \emph{edit distance}~\cite{Masek-Paterson}. Can edit distance be computed in $n^2/2^{(\log n)^{\delta}}$ time?

{\bf 3. Practicality?} There are two potential impediments to making the approach of this paper work in practice: (1) the translation from $\AC^0[2]$ circuits to polynomials, and (2) Coppersmith's matrix multiplication algorithm. For case (1), there are no large hidden constants inherent in the Razborov-Smolensky translation, however the expansion of the polynomial as an XOR of ANDs yields a quasi-polynomial blowup. A careful study of alternative translations into polynomials would likely improve this step for practice. For case (2), Coppersmith's algorithm as described in Appendix~\ref{coppersmith-appendix} consists of a series of multiplications with Vandermonde and inverse Vandermonde matrices (which are very efficient), along with a recursive step on $2 \times 3$ and $3 \times 2$ matrices, analogous to Strassen's famous algorithm. We see no theoretical reason why this algorithm (implemented properly) would perform poorly in practice, given that Strassen's algorithm can be tuned for practical gains~\cite{Grayson96,CLPT02,DN09,Ballard12,Ballard12comm}. Nevertheless, it would likely be a substantial engineering challenge to turn the algorithms of this paper into high-performance software.

{\bf 4. APSP For Sparse Graphs?} Perhaps a similar approach could yield an APSP algorithm for $m$-edge, $n$-node graphs running in $\tilde{O}(mn/2^{(\log n)^{\delta}}+n^2)$ time, which is open even for undirected, unweighted graphs. (The best known algorithms are due to Chan~\cite{Chan12} and take roughly $mn/\log n$ time.)

{\bf 5. Truly Subcubic APSP?} What other circuit classes can compute $(\min,+)$ inner product and also permit a fast evaluation algorithm on many inputs? This question now appears to be central to the pursuit of truly subcubic ($n^{3-\eps}$ time) APSP. Although we observe in the paper that $(\min,+)$ inner product is efficiently computable in $\AC^0$, the usual algebraic $(+,\times)$ inner product is in fact \emph{not} in $\AC^0$. (Multiplication is not in $\AC^0$, by a reduction from Parity~\cite{ChandraStockmeyerVishkin84}.) This raises the intriguing possibility that $(\min,+)$ matrix product (and hence APSP) is not only in truly subcubic time, but could be \emph{easier} than integer matrix multiplication. A prerequisite to this possibility would be to find new Boolean matrix multiplication algorithms which do not follow the Strassenesque approaches of the last 40+ years. Only minor progress on such algorithms has been recently made~\cite{BansalWilliams}.

\paragraph{Acknowledgements.} Many thanks to Virginia Vassilevska Williams; without her as a sounding board, I would have stopped thinking about APSP a long time ago. I'm also grateful to Uri Zwick, Farah Charab, and the STOC referees for helpful comments. Thanks also to Eric Allender for a discussion on references.

\bibliographystyle{alpha}
\bibliography{apsp}

\appendix

\section{Proof of Lemma~\ref{AC0minplus}}\label{AC0minplus-appendix}

\begin{reminder}{Theorem~\ref{AC0minplus}} Given $u,v \in (([0,M] \cap \Z)\cup\{\infty\})^d$ encoded as $O(d \log M)$-bit strings, $(u \star v)$ is computable with constant-depth AND/OR/NOT circuits of size $(d \log M)^{O(1)}$. That is, the min-plus inner product function is computable in $\AC^0$ for every  $d$ and $M$.
\end{reminder}

\begin{proof} Before giving the circuit, let us briefly discuss the encoding of $\infty$. In principle, all we need is that $\infty$ encodes an integer greater than $2M$, and that addition of $\infty$ with any number equals $\infty$ again. With that in mind, we use the following convention: $t=3+\log M$ bits are allocated to encode each number in $\{0,\ldots,M\}$ (with two leading zeroes), and $\infty$ is defined as the all-ones string of $t$ bits.

The addition of two $t$-bit strings $x$ and $y$ is computable in $t^{O(1)}$ size and constant depth, using a ``carry-lookahead'' adder~\cite{StockmeyerV84} (with an improved size bound in~\cite{ChandraFL85}). Recall that a carry-lookahead adder determines in advance: \begin{compactitem}
\item which bits of $x$ and $y$ will ``generate'' a $1$ when added, by taking the AND of each matching pair of bits from $x$ and $y$, and
\item which bits of $x$ and $y$ will ``propagate'' a $1$ if given a $1$ from the summation of lower bits, by taking the XOR of each matching pair of bits.
\end{compactitem}
The ``generate'' and ``propagate'' bits can be generated in constant depth. Given them, in parallel we can determine (for all $i$) which bits $i$ of the addition will generate a carry in constant depth, and hence determine the sum in constant depth. (To handle the case of $\infty$, we simply add a side circuit which computes in parallel if one of the inputs is all-$1$s, in which case all output bits are forced to be $1$.)

Chandra, Stockmeyer, and Vishkin~\cite{ChandraStockmeyerVishkin84} show how to compute the minimum of a collection of numbers (given as bit strings) in $\AC^0$. For completeness, we give a construction here. First, the comparison of two $t$-bit strings $x$ and $y$ (construed as non-negative integers) is computable in $\AC^0$. Define
\[LEQ(x,y) = \left(\bigwedge_{i=1}^{t} (1 + x_i + y_i)\right) \vee \bigvee_{i=1}^{t} \left((1 + x_i) \wedge y_i \wedge \bigwedge_{j=1}^{i-1} (1 + x_j + y_j)\right),\] where $+$ stands for addition modulo $2$. The first disjunct is true if and only if $x=y$. For the second disjunct and given $i=1,\ldots,t$, the inner expression is true if and only if the first $i-1$ bits of $x$ and $y$ are equal, the $i$th bit of $x$ is $0$, and the $i$th bit of $y$ is 1. Replacing each $1+a+b$ with $(\neg a \vee  b)\wedge(a \vee \neg b)$, we obtain an $\AC^0$ circuit.

To show that the minimum of $d$ $t$-bit strings $x_1,\ldots,x_d$ is in $\AC^0$, we first prove it for the case where the minimum is \emph{unique} (in the final algorithm of Theorem~\ref{main}, this will be the case). The expression $MIN(x_i) := \bigwedge_{~j \in \{1,\ldots,d\}} LEQ(x_i,x_j)$ is true if and only if $x_i = \min_j x_j$. When the  minimum is unique, the following computes the $\ell$th bit of the minimum $x_i$: \[MIN_{\ell}(x_1,\ldots,x_d) := \bigvee_{\substack{i = 1,\ldots,d \\\text{$\ell$th bit of $i$ is $1$}}} MIN(x_i).\]

Finally, let us handle the case where there is more than one minimum $x_i$. We will compute the minimum $i^{\star}$ such that $MIN(x_{i^{\star}})=1$, then output the $\ell$th bit of $x_{i^{\star}}$. Define $MINBIT(x_i,i,j)$ to be true if and only if ($MIN(x_i)$ and the $j$th bit of $i$ is $1$) or $\neg MIN(x_i)$. It is easy to see that $MINBIT$ is in $\AC^0$. For all $i=1,\ldots,d$, compute the $d$-bit string $f(x_i) := MINBIT(x_i,i,1)\cdots MINBIT(x_i,i,d)$ in constant depth. The function $f$ maps every non-minimum string $x_{i'}$ to the all-ones $d$-bit string, and each minimum string $x_i$ to its position $i$ as it appeared in the input. Now, computing the minimum over all $f(x_i)$ determines the smallest $i^{\star}$ such that $x_{i^{\star}}$ is a minimum; this minimum can be found in constant depth, as observed above.
\end{proof}

\section{Appendix: Derandomizing the APSP algorithm}

\label{detAPSP-appendix}

\begin{reminder}{Theorem~\ref{detAPSP}} There is a $\delta > 0$ and a deterministic algorithm for APSP running in $n^3/2^{(\log n)^{\delta}}$ time on the real RAM.
\end{reminder}

The proof combines the use of Fredman's trick in Theorem~\ref{main} with the deterministic reduction from circuits to polynomials in Theorem~\ref{simpleAPSP}. 

Take the $\AC^0[2]$ circuit $C$ for computing min-plus inner products on length $d$ vectors, as described in the proof of Theorem~\ref{main} of Section~\ref{mainresult}. The circuit $C$ is comprised of circuits $C_1,\ldots,C_{\log d}$ such that each $C_{\ell}$ takes a bit string of length $k=2d^2\log(2n)$ representing two vectors $u,v$ from $\{1,\ldots,2n\}^d$, and outputs the $\ell$th bit of the smallest $k^{\star}$ such that the min-plus inner product of $u$ and $v$ equals $u[k^{\star}]+v[k^{\star}]$.

Applying Lemma~\ref{Beigel-Tarui} from Theorem~\ref{simpleAPSP}, we can reduce each $C_{\ell}$ to a $\SYM$ circuit $D_{\ell}$ of size $2^{\log^c k}$ for some constant $c \geq 1$. Then, analogously to Theorem~\ref{polyeval}, we reduce the evaluation of $D_{\ell}$ on inputs of length $k$ to an inner product (over $\Z$) of two $0-1$ vectors $u',v'$ of length $2^{\log^c k}$. For every AND gate $g$ in $D$ that is an AND of bits $\{x_{i_1},\ldots,x_{i_t},\ldots,y_{j_1},\ldots,y_{j_t}\}$, we associate it with a component $g$ in the two vectors; in the $g$th component of $u'$ we multiply all the $x_{i_k}$ bits owned by $u$, and in the $g$th component of $v'$ we multiply all the $y_{j_k}$ bits owned by $v$.

Therefore we can reduce an $n \times d$ and $d \times n$ min-plus matrix product to a matrix product over the integers, by replacing each row $u$ of the first matrix by a corresponding $u'$ of length \[\ell = 2^{\log^c k} = 2^{(\log(2d^2\log(2n)))^c} \leq 2^{(3\log d + 2\log \log n)^c},\] and replacing each column $v$ of the second matrix by a corresponding $v'$ of length $\ell$. When $d$ is small  enough to satisfy $2^{(3\log d + 2\log \log n)^c} \leq n^{0.1}$, this is a reduction from $n \times d$ and $d \times n$ min-plus product to $n \times n^{0.1}$ and $n^{0.1} \times n$ matrix product over $\Z$, with matrices containing $0$-$1$ entries. As argued earlier, this implies an $\tilde{O}(n^3/d)$ time algorithm for min-plus product.

We derive an upper bound on $d$ as follows:
\begin{eqnarray*} 2^{(3\log d + 2\log \log n)^c} &\leq& n^{0.1} \\
\iff 3\log d + 2\log \log n &\leq& (0.1 \log n)^{1/c}\\
\iff \log d &\leq& \frac{(0.1 \log n)^{1/c}-2\log \log n}{3},
\end{eqnarray*}
hence $d = 2^{(0.1 \log n)^{1/c}/4}$ suffices for sufficiently large $n$.

Examining the proof of Theorem~\ref{Beigel-Tarui} shows that we can estimate $c = 2^{\Theta(d')}$, where $d'$ is the depth of the original $\AC^0[2]$ circuit. However, as Beigel and Tarui's proof also works for the much more expressive class $\ACC$ (and not just $\AC^0[2]$), we are confident that better estimates on $c$ are possible with a different argument, and hence refrain from calculating an explicit bound here.

\section{Appendix: An exposition of Coppersmith's algorithm}
\label{coppersmith-appendix}

In 1982, Don Coppersmith proved that the rank (that is, the number of essential multiplications) of $N \times N^{0.1}$ and $N^{0.1} \times N$ matrix multiplication is at most $O(N \log^2 N)$. Prior work has observed that his algorithm can also be used to show that the total number of arithmetic operations for the same matrix multiply is $N \cdot \poly(\log N)$. However, the implication is not immediate, and uses specific properties of Coppersmith's algorithm. Because this result is so essential to this work and another recent circuit lower bound~\cite{Williams13thr}, we give a self-contained exposition here.

\begin{theorem}[Coppersmith~\cite{Coppersmith82}]\label{rank} For all sufficiently large $N$, the rank of $N \times N^{.1} \times  \times N$ matrix multiplication is at most $O(N^2 \log^2 N)$. \end{theorem}

We wish to derive the following consequence of Coppersmith's construction, which has been mentioned in the literature before~\cite{Seroussi,ApplebaumCPS09,Williams11}:

\begin{lemma}\label{coppersmith} For all sufficiently large $N$, and $\alpha \leq .172$, multiplication of an $N \times N^{\alpha}$ matrix with an $N^{\alpha} \times N$ matrix can be done in $N^2 \cdot \poly(\log N)$ arithmetic operations, over any field with $O(2^{\poly(\log N)})$ elements.\end{lemma}

For brevity, we will use the notation ``$\ell \times m \times n$ matrix multiply'' to refer to the multiplication of $\ell \times m$ and $m \times n$ matrices (hence the above gives an algorithm for $N \times N^{\alpha} \times N$ matrix multiply).

Note Lemma~\ref{coppersmith} has been ``improved'' in the sense that the upper bound on $\alpha$ has been increased mildly over the years~\cite{Coppersmith97,HP98,Ke08,LeGall12}. However, these later developments only run in $N^{2+o(1)}$ time, not $N^2 \cdot \poly(\log N)$ time (which we require). Our exposition will expand on the informal description given in recent work~\cite{Williams11}.

First, observe that the implication from Theorem~\ref{rank} to Lemma~\ref{coppersmith} is not immediate. For example, it could be that Coppersmith's algorithm is non-uniform, making it difficult to apply. As far as we know, one cannot simply take ``constant size'' arithmetic circuits implementing the algorithm of Theorem~\ref{rank} and recursively apply them. In that case, the $\poly(\log N)$ factor in the running time would then become $N^{\eps}$ for some constant $\eps > 0$ (depending on the size of the constant-size circuit). To keep the overhead polylogarithmic, we have to unpack the algorithm and analyze it directly.

\subsection{A short preliminary}

Coppersmith's algorithm builds on many other tools from prior matrix multiplication algorithms, many of which can be found in the highly readable book of Pan~\cite{Pan}. Here we will give a very brief tutorial of some of the aspects.

\paragraph{Bilinear algorithms and trilinear forms.} Essentially all methods for matrix multiplication are bilinear (and if not, they can be converted into such algorithms), meaning that they can be expressed in the so-called trilinear form
\begin{equation}\label{trilinear} \sum_{ijk} A_{ik}B_{kj}C_{ji} + p(x) = \sum_{\ell=1}^5 (\sum_{ij} \alpha_{ij} A_{ij})\cdot (\sum_{ij} \beta_{ij} B_{ij})\cdot(\sum_{ij} \gamma_{ij} C_{ij})\end{equation}
where $\alpha_{ij}$, $\beta_{ij}$, and $\gamma_{ij}$ are constant-degree polynomials in $x$ over the field, and $p(x)$ is a polynomial with constant coefficient $0$. Such an algorithm can be converted into one with no polynomials and minimal extra overhead (as described in Coppersmith's paper). Typically one thinks of $A_{ik}$ and $B_{kj}$ as entries in the input matrices, and $C_{ji}$ as indeterminates, so the LHS of~\eqref{trilinear} corresponds to a polynomial whose $C_{ji}$ coefficient is the $ij$ entry of the matrix product. Note the {\em transpose} of the third matrix $C$ corresponds to the final matrix product.

To give an explicit example, we assume the reader is familiar with Strassen's famous method for $2 \times 2 \times 2$ matrix multiply. Strassen's algorithm can be expressed in the form of~\eqref{trilinear} as follows:
\begin{eqnarray}\label{strassen}
\sum_{i,j,k=0,1} A_{ik}B_{kj}C_{ji} &=& (A_{00} + A_{11})(B_{00} + B_{11})(C_{00} + C_{11})\\ \nonumber
& & + (A_{10} + A_{11})B_{00}(C_{01}-C_{11}) + A_{00}(B_{01} - B_{11})(C_{10} + C_{11})\\ \nonumber
& & + (A_{10} - A_{00})(B_{00} + B_{01})C_{11} + (A_{00} + A_{01})B_{11}(C_{10} - C_{00}) \\ \nonumber
& & + A_{11}(B_{10} - B_{00})(C_{00} + C_{01}) + (A_{01} - A_{11})(B_{10}+B_{11})C_{00}. \nonumber\end{eqnarray}
The LHS of~\eqref{trilinear} and~\eqref{strassen} represents the trace of the product of three matrices $A$, $B$, and $C$ (where the $ij$ entry of matrix $X$ is $X_{ij}$). It is well known that every bilinear algorithm naturally expresses multiple algorithms through this trace representation. Since \[tr(ABC) = tr(BCA) = tr(CAB)=tr((ABC)^T)=tr((BCA)^T)=tr((CAB)^T),\] if we think of $A$ as a symbolic matrix and consider~\eqref{trilinear}, we obtain a new algorithm for computing a matrix $A$ when given $B$ and $C$. Similarly, we get an algorithm for computing a $B$ when given $A$ and $C$, and analogous statements hold for computing $A^T$, $B^T$, and $C^T$. So the aforementioned algorithm for multiplying a sparse $2 \times 3$ and sparse $3 \times 2$ yields several other algorithms.

\paragraph{Sch\"{o}nhage's decomposition paradigm.} Coppersmith's algorithm follows a specific paradigm introduced by Sch\"{o}nhage~\cite{Schoenhage} which reduces arbitrary matrix products to slightly larger matrix products with ``structured nonzeroes.'' The general paradigm has the following form. Suppose we wish to multiply two matrices $A''$ and $B''$.
\begin{compactenum}
\item  First we {\em preprocess} $A''$ and $B''$ in some efficient way, decomposing $A''$ and $B''$ into structured matrices $A,A',B,B'$ so that $A'' \cdot B'' = A' \cdot A \cdot B \cdot B'$. (Note, the dimensions of $A' \cdot A$ may differ from $A''$, and similarly for $B' \cdot B$ and $B''$.) The matrices $A$ and $B$ are sparse ``partial'' matrices directly based on $A''$ and $B''$, but they have larger dimensions, and only contain nonzeroes in certain structured parts. The matrices $A'$ and $B'$ are very simple and explicit matrices of scalar constants, chosen independently of $A''$ and $B''$. (In particular, $A'$ and $B'$ are Vandermonde-style matrices.)

\item Next, we apply a specialized constant-sized matrix multiplication algorithm in a recursive manner, to multiply the structured $A$ and $B$ essentially optimally. Recall that Strassen's famous matrix multiplication algorithm has an analogous form: it starts with a seven-multiplication product for $2 \times 2 \times 2$ matrix multiplication, and recursively applies this to obtain a general algorithm for $2^M \times 2^M \times 2^M$ matrix multiplication. Here, we will use an \emph{optimal} algorithm for multiplying constant-sized matrices with zeroes in some of the entries; when this algorithm is recursively applied, it can multiply sparse $A$ and $B$ with nonzeroes in certain structured locations.

\item Finally, we {\em postprocess} the resulting product $C$ to obtain our desired product $A'' \cdot B''$, by computing $A' \cdot C \cdot B'$. Using the simple structure of $A'$ and $B'$, the matrix products $D := A' \cdot C$ and $D \cdot B'$ can be performed very efficiently. Our aim is to verify that each step of this process can be efficiently computed, for Coppersmith's full matrix multiplication algorithm.

\end{compactenum}

\subsection{The algorithm}

The construction of Coppersmith begins by taking input matrices $A''$ of dimensions $2^{4M/5} \times {M \choose 4M/5}2^{4M/5}$ and $B''$ of dimensions ${M \choose 4M/5}2^{4M/5} \times 2^{M/5}$ where $M \approx \log N$, and obtains an $O(5^M \poly(M))$ algorithm for their multiplication. Later, he symmetrizes the construction to get an $N \times N \times N^{\alpha}$ matrix multiply. We will give this starting construction and show how standard techniques can be used to obtain an $N \times N^{\alpha} \times N$ matrix multiply from his basic construction.

The multiplication of $A''$ and $B''$ will be derived from an algorithm which computes the product of $2 \times 3$ and $3 \times 2$ matrices with zeroes in some entries. In particular the matrices have the form:
\[\left( \begin{array}{ccc} a_{11} & a_{12} & a_{13} \\ 0 & a_{22} & a_{23} \end{array} \right), \left( \begin{array}{cc} b_{11} & b_{12}\\ b_{21} & 0 \\ b_{31} & 0 \end{array} \right), \] and the algorithm is given by the trilinear form \begin{eqnarray}\label{2332}(a_{11} + x^2 a_{12})(b_{21} + x^2 b_{11})(c_{11}) + (a_{11} +x^2 a_{13}(b_{31})(c_{11} - x c_{21}) + (a_{11} + x^2 a_{22})(b_{21} - x b_{21})(c_{22})\\ \nonumber + (a_{11}+x^2 a_{23})(b_{31} + x b_{12}) (c_{12} + x c_{21}) - (a_{11})(b_{21}+b_{31})(c_{11}+c_{12})\\ \nonumber = x^2(a_{11}b_{11}c_{11} + a_{11}b_{12}c_{21} + a_{12}b_{21}c_{11} + a_{13}b_{31}c_{11} + a_{22}b_{21}c_{12} + a_{23}b_{31}c_{12}) + x^3 \cdot P(a,b,c,x).\end{eqnarray} That is, by performing the five products of the linear forms of $a_{ij}$ and $b_{k\ell}$ on the LHS, and using the $c_{ij}$ to determine how to add and subtract these products to obtain the output $2 \times 2$ matrix, we obtain a polynomial in each matrix entry whose $x^2$ coefficients yield the final matrix product $c_{ij}$.

When the algorithm given by \eqref{2332} is applied recursively to $2^M \times 3^M$ and $3^M \times 2^M$ matrices (analogously to how Strassen's algorithm is applied to do $2^M \times 2^M \times 2^M$ matrix multiply), we obtain an algorithm that can multiply matrices $A$ and $B$ with dimensions $2^M \times 3^M$ and $3^M \times 2^M$, respectively, where $A$ has $O(5^M)$ nonzeroes, $B$ has $O(4^M)$ nonzeroes, and these nonzeroes appear in a highly regular pattern (which can be easily deduced). This recursive application of \eqref{2332} will result in polynomials in $x$ of degree $O(M)$, and additions and multiplications on such polynomials increase the overall time by an $M \cdot \poly(\log M)$ factor. Therefore we can multiply these $A$ and $B$ with structured nonzeroes in $O(5^M \cdot \poly(M))$ field operations.

The decomposition of $A''$ and $B''$ is performed as follows. We choose $A'$ and $B'$ to have dimensions $2^{4M/5} \times 2^{M}$ and $2^M \times 2^{M/5}$, respectively, and such that  all $2^{4M/5} \times 2^{4M/5}$ submatrices of $A'$ and $2^{M/5} \times 2^{M/5}$ submatrices of $B'$ are non-singular. Following Sch\"{o}nhage, we pick $A'$ and $B'$ to be rectangular Vandermonde matrices: the $i,j$ entry of $A'$ is $(\alpha_j)^{i-1}$, where $\alpha_1,\alpha_2,\ldots$ are distinct elements of the field; $B'$ is defined analogously. Such matrices have three major advantages: (1) they can be succinctly described (with $O(2^M)$ field elements), (2) multiplying these matrices with arbitrary vectors can be done extremely efficiently, and (3) inverting an arbitrary square submatrix can be done extremely efficiently. More precisely, $n\times n$ Vandermonde matrices can be multiplied with arbitrary $n$-vectors in $O(n \cdot \poly(\log n))$ operations, and computing the inverse of an $n \times n$ Vandermonde matrix can be done in $O(n \cdot \poly(\log n))$ operations (for references, see~\cite{CKY,BiniPan}). In general, operations on Vandermonde matrices, their transposes, their inverses, and the transposes of inverses can be reduced to fast multipoint computations on univariate polynomials. For example, multiplying an $n\times n$ Vandermonde matrix with a vector is equivalent to evaluating a polynomial (with coefficients given by the vector) on the $n$ elements that comprise the Vandermonde matrix, which takes $O(n \log n)$ operations. This translates to $O(n \cdot \poly(\log n))$ arithmetic operations.

The matrices $A$ and $B$ have dimensions $2^M \times 3^M$ and $3^M \times 2^M$, respectively, where $A$ has  only $O(5^M)$ nonzeroes, $B$ has only $O(4^M)$ nonzeroes, and there is an optimal algorithm for multiplying $2 \times 3$ (with 5 nonzeroes) and $3 \times 2$ matrices (with 4 nonzeroes) that can be recursively applied to multiply $A$ and $B$ optimally, in $O(5^M \cdot \poly(M))$ operations. Matrices $A$ and $B$ are constructed as follows: take any one-to-one mapping between the ${M \choose 4M/5}2^{M/5}$ columns of the input $A''$ and columns of the sparse $A$ with exactly $2^{4M/5}$ nonzeroes. For these columns $q$ of $A$ with $2^{4M/5}$ nonzeroes, we compute the inverse $A_q^{-1}$ of the $2^{4M/5} \times 2^{4M/5}$ minor $A_q$ of $A'$ with rows corresponding to the nonzeroes in the column, and multiply $A_q^{-1}$ with column $q$ (in $2^{4M/5} \cdot \poly(M)$ time). After these columns are processed, the rest of $A$ is zeroed out. Then, there is a one-to-one correspondence between columns of $A''$ and nonzero columns of $A' \cdot A$. Performing a symmetric procedure for $B''$ (with the same mapping on rows instead of columns), we can decompose it into $B$ and $B'$ such that there is a one-to-one correspondence between rows of $B''$ and nonzero rows of $B\cdot B'$. It follows that this decomposition takes only $O({M \choose 4M/5}2^{4M/5} \cdot 2^{4M/5} \cdot \poly(M))$ time. Since $5^M \approx {M \choose 4M/5}4^{4M/5}$ (within $\poly(M)$ factors), this quantity is upper bounded by $5^M \cdot \poly(M)$.

After $A$ and $B$ are constructed, the constant-sized algorithm for $2 \times 3$ and $3 \times 2$ mentioned above can be applied in the usual recursive way to multiply the sparse $A$ and $B$ in $O(5^M \cdot \poly(M))$ operations; call this matrix $Z$. Because $A'$ and $B'$ are Vandermonde, the product $A' \cdot Z \cdot B'$ can be computed in $O(5^M \cdot \poly(M))$ operations. Hence we have an algorithm for multiplying matrices of dimensions $2^{4M/5} \times {M \choose 4M/5}2^{4M/5}$ and ${M \choose 4M/5}2^{4M/5} \times 2^{M/5}$ that is explicit and takes $5^M \cdot \poly(M)$ operations.

Call the above algorithm {\sc Algorithm 1}. Observe {\sc Algorithm 1} also works when the entries of $A''$ and $B''$ are themselves matrices over the field. (The running time will surely increase in proportion to the sizes of the underlying matrices, but the bound on the number of {\em operations on the entries} remains the same.)

Up to this point, we have simulated Coppersmith's construction completely, and have simply highlighted its efficiency. By exploiting the symmetries of matrix multiplication algorithms in a standard way, we can extract more algorithms from the construction. The trace identity tells us that \[tr(ABC) = tr(BCA),\] implying that the expression \eqref{2332} can also be used to partially multiply a $3^M \times 2^M$ matrix $B$ with at most $4^M$ structured nonzeroes and ``full'' $2^M \times 2^M$ matrix $C$ in $5^M \cdot \poly(M)$ operations, obtaining a $3^M \times 2^M$ matrix $A^T$ with at most $5^M$ nonzeroes. In our {\sc Algorithm 1}, we have a decomposition of $A$ and $B$; in terms of the trace, we can derive: \[tr(A'' B'' \cdot C'') = tr(A'A \cdot B B' \cdot C'') = tr(B \cdot B' C'' A' \cdot A).\]

This can be applied to obtain an algorithm for ${M \choose 4M/5}2^{4M/5} \times 2^{M/5} \times 2^{4M/5}$ matrix multiplication, as follows. Given input matrices $B''$ and $C''$ of the respective dimensions, we decompose $B''$ into a $3^M \times 2^M$ $B$ with $O(4^M)$ nonzeroes and $2^{M} \times 2^{M/5}$ Vandermonde $B'$, as described above. Letting $A'$ be a Vandermonde $2^{4M/5} \times 2^{M}$ matrix, we compute the matrix $C := B' \cdot C'' \cdot A'$ in at most $4^M \cdot \poly(M)$ operations. Noting that $C$ is $2^M \times 2^M$, we can then multiply $B$ and $C$ in $5^M \cdot \poly(M)$ operations. This results in a $3^M \times 2^M$ matrix $A^T$ with at most $5^M$ nonzeroes. The final output $A''$ is obtained by using the one-to-one mapping to extract the appropriate ${M \choose 4M/5}2^{4M/5}$ rows from $A^T$, and multiplying each such row by the appropriate inverse minor of $A'$ (corresponding to the nonzeroes of that row). This takes at most ${M \choose 4M/5}2^{4M/5} \cdot 2^M \cdot \poly(M) \leq 5^M \cdot \poly(M)$ operations. Call this {\sc Algorithm 2}.

From {\sc Algorithm 2} we immediately obtain an algorithm for $2^{4M/5} \times 2^{M/5} \times {M \choose 4M/5}2^{4M/5}$ matrix multiplication as well: given input matrices $(C'')^T$ and $(B'')^T$ of te respective dimensions, simply compute $B'' \cdot C''$ using {\sc Algorithm 2}, and output the transpose of the answer. Call this {\sc Algorithm 3}.

Finally, by ``tensoring'' {\sc Algorithm 2} with {\sc Algorithm 3}, we derive an algorithm for matrix multiplication with dimensions \[{M \choose 4M/5}2^{4M/5}\cdot 2^{4M/5} \times 2^{2M/5} \times {M \choose 4M/5}2^{4M/5}\cdot 2^{4M/5} \geq 5^M/M \times 4^{M/5} \times 5^M/M.\]

That is, we divide the two input matrices of large dimensions into blocks of $2^{4M/5} \times 2^{M/5}$ and $2^{M/5} \times {M \choose 4M/5}2^{4M/5}$ dimensions, respectively. We execute {\sc Algorithm 2} on the blocks, and call {\sc Algorithm 3} when the product of two blocks is needed.

As both {\sc Algorithm 2} and {\sc Algorithm 3} are explicit and efficient, their ``tensorization'' inherits these properties. {\sc Algorithm 2} uses $5^M \cdot \poly(M)$ operations, and each operation can take up to $5^M \cdot \poly(M)$ time (due to calls to {\sc Algorithm 3}). Therefore, we can perform a $5^M \times 4^{2M/5} \times 5^M$ matrix multiply over fields with $2^{\poly(M)}$ elements, in $5^{2M} \cdot \poly(M)$ time. Setting $n = \log(M)/\log(5)$, the algorithm runs in $n^2 \cdot \poly(\log n)$ time for fields with $2^{\poly(\log n)}$ elements.

\end{document}